\documentstyle[prc,aps,epsf,psfig,amssymb]{revtex}

\newcommand \be{\begin{eqnarray}}
\newcommand \ee{\end{eqnarray}}
\tightenlines
\begin{document}
\title{Low Momentum NN-interactions and Nuclear Matter
calculations.}
\author{H. S. K\"ohler\\}
\address{Physics Department, University of Arizona, Tucson, Arizona
85721}
\date{\today}
\maketitle
\begin{abstract}
Separable  nucleon-nucleon potentials are calculated using inverse
scattering techniques as presented in previously published work.
The dependence of the potentials on the
momentum cut-off of the scattering phase-shifts is studied. 
Some comparison is made with the $V_{low-k}$ potential.
The effect of the cut-off on nuclear matter binding energy calculated by
standard Brueckner theory is also presented. 
It is found that a cut-off larger than about 4 $fm^{-1}$ will keep the
error to within one $MeV$ around saturation density. 
While the potentials are cut-off dependent
the effective interaction represented by the Brueckner $K$-matrix
is less sensitive to this cut-off. This is in particular found to be the
case for the $^{1}S_{0}$ state.

\end{abstract}

\section{Introduction}
EFT methods promise to provide  information on NN-interactions for
studies of low energy nuclear systems. Traditional methods used to
obtain NN-potentials consist of a mixture of contributions from
meson-exchange forces and phenomenological adjustments. The OPEP part of
the potential is fairly well established while the main uncertainty
concerns the short-ranged repulsion. The philosophy behind the EFT
approach is that low-energy nuclear phenomena should be independent of
the details of the short-ranged repulsions which are of a high energy
origin and can be integrated out.  This is somewhat akin to the features 
of the Moszkowski-Scott separation method \cite{mos60} where it is
explicitly shown that 
although the short-ranged correlations associated
with the short-ranged repulsions are to a large degree responsible for
the saturation of nuclear matter. The details of the repulsions are
however not 
important, but are absorbed by an integrated quantity referred to as 
the "wound-integral".
It provides for a density-dependent repulsion in the in-medium effective
interaction to be supplemented with the long-ranged part that the method
prescribes.

Related to the above one may ask about the relation between nuclear
matter binding/saturation  and the NN-repulsions as exhibited by the
scattering phase-shifts of high momenta. 
Some scattered information is available from
the more than 50 years of publications on nuclear matter calculations.
It seems however appropriate to (re)examine this relationship by
studying the effect of nuclear matter binding/staturation on the
momentum cutoff
of the scattering phaseshifts entering the construction of the
NN-potentials.

Because the $S$ states have poles near $E=0$ a reasonable ansatz is to
assume that these potentials are separable.\cite{bro76}
In a previous publication \cite{kwo95} a separable potential for all
states considered (21) was
calculated using established inverse scattering techniques. 
The input
was experimental scattering phase-shifts and deuteron data.  This does
of course not provide a unique solution. 
One important aspect of this work
is however that the phase-shifts are fitted EXACTLY. 
The Arndt set of
phase-shifts was chosen \cite{arn87}. 
These provide on-shell
data while off-shell parts of the interactions are constrained
essentially by the chosen rank of the potentials.
The rank of
the potentials in the different states was kept as low as was possible
while  complying with the experimental input. By increasing the rank the
off-diagonal parts can be modified as found suitable, but not much
experimental information is available here. As already pointed out by
Tabakin \cite{tab69}
the off-diagonal parts are also affected
by the chosen momentum (energy) cut-off.
The binding energy of nuclear matter was calculated by standard
Brueckner techniques.
Comparison with calculations using the Bonn-potential \cite{mac89}
showed
excellent agreement  except in the $^{3}P_{1}$ state. In this state the
half-shell reactance matrix elements showed a large difference from
those of ref.\cite{mac89}. 

This method of constructing the NN-potential is very suitable for the
present study of momentum cut-offs, because of the ease of the calculations
while maintaing exact fits to experimental data.
The Arndt phases are defined up to 1.6 GeV lab energy which converts to
$k \approx 4.3 fm^{-1}$ in the C.M. system. In the previous work the
difference between some extrapolations up to $10 fm^{-1}$ was
investigated. It was found although that the resulting potentials
changed appreciably there was little or no change in the corresponding
(half-shell) reactance matrices for momenta relevant to low energy
nuclear physics such as nuclear saturation. 

This finding is explored here in some more detail showing  nuclear
matter bindings and saturation curves for cutoffs as low as at $2 fm^{-1}$. 
The corresponding potential parameters are also shown as a function of
cut-offs and compared with $V_{low-k}$.

\section{Numerical results}
All calculations were done essentially as in the previous
work.\cite{kwo95} 
It may be of interest that the time to
compute the separable potentials from the 21 phaseshifts AND a 
many-body Brueckner self-consistent calculation at a fixed density
only takes seconds which is one reason why an investigation of this type
is feasible.
\subsection{The $^{1}S_{0}$ potential and K-matrix}

\begin{figure}
\centerline{
\psfig{figure=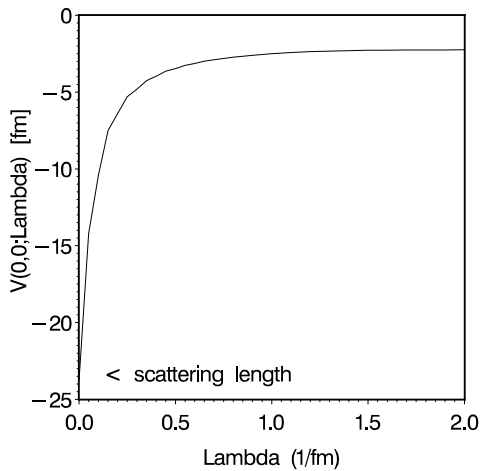,width=7cm,angle=0}
}
\vspace{.0in}
\caption{
$V(k=0,k=0)$ as a function of the cutoff $\Lambda$.
}
\label{vlowfig6}
\end{figure}

The separable potential is calculated for different values of the
cut-off momentum of the phase-shifts. The potential is consequently only
defined for momenta smaller than (or equal to) the chosen cut-off.
If the cutoff-momentum given in the C.M. system is larger than the
maximum momentum $4.3 fm^{-1}$ of the Arndt phases they are extrapolated
by a straight line to the chosen cut-off momentum. 
Because one incentive for this investigation has 
been the development of the low
momentum NN-interaction referred to as $V_{low \ k}$
\cite{bog01,bog03,sch03}
Fig \ref{vlowfig6} shows $V(k=0,k=0)$ as a function of the cutoff
$\Lambda$ calculated by inverse scattering from the $^{1}S_{0}$
phase-shifts.
One finds that it is practically identical to the result shown in
published papers on $V_{low \ k}$ . At the limit $\Lambda \Rightarrow 0$
one finds $V(0,0)=-23.73$ i.e. the scattering length.

\begin{figure}
\centerline{
\psfig{figure=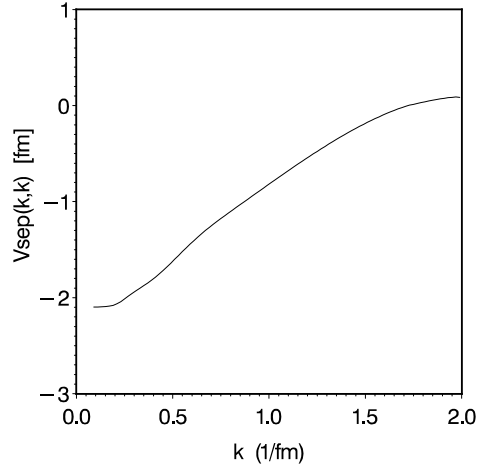,width=7cm,angle=0}
}
\vspace{.0in}
\caption{
The separable potential $V(k,k)$ in the $^{1}S_{0}$ state
for a cut-off $\Lambda=2.$.
}
\label{vsep1s1}
\end{figure}

A plot of the diagonal elements  of the same separable potential $V(k,k)$
with $\Lambda=2.$ is shown by Fig \ref{vsep1s1}. 
This also shows almost perfect agreement with the cited publications.
It is probably not necessary to point out the completely different
method used here as compared to that of the $V_{low \ k}$ derivation.
The only similarity is really that both rely on fit to experimental
scattering data; which in our case is exact. 

The above results serve as a comparison with $V_{low \ k}$. However the
main purpose here is to investigate in more detail the relation between
cut-offs $\Lambda$ and the potential as well as the in-medium
interaction (The Brueckner K-matrix) and nuclear binding.

\begin{figure}
\centerline{
\psfig{figure=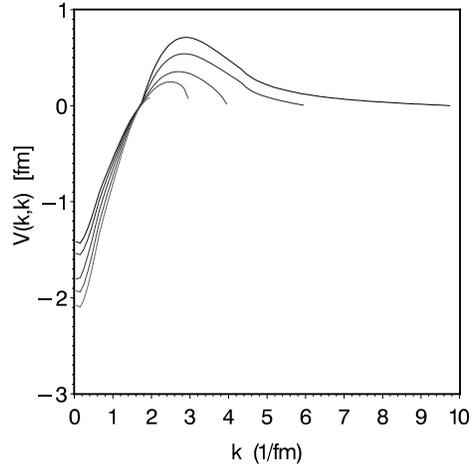,width=7cm,angle=0}
}
\vspace{.0in}
\caption{
The diagonal elements of the separable $^{1}S_{0}$ potential 
as a function of momentum $k$ for different values of cut-off
$\Lambda$. Each curve ends at its value of $\Lambda$.
}
\label{Vkk}
\end{figure}

Fig \ref{Vkk} shows the dependence of the diagonal parts of the $^{1}S_{0}$ 
potential. One finds a rather strong dependence.

\begin{figure}
\centerline{
\psfig{figure=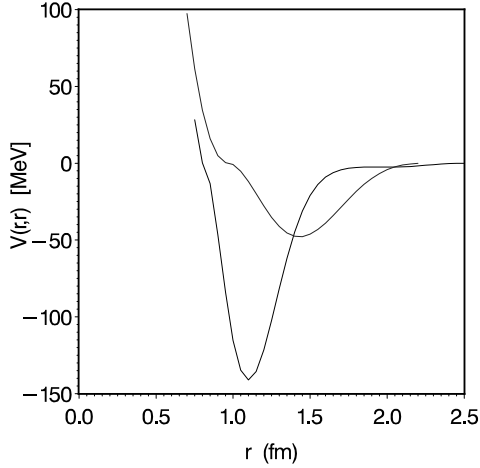,width=7cm,angle=0}
}
\vspace{.0in}
\caption{
Separable $^{1}S_{0}$ potential in coordinate space for $\Lambda=6.$
(lower curve) and for $\Lambda=4.$ (upper curve).
}
\label{vsepr}
\end{figure}

The dependence on cut-off is even more striking in ccordinate space as
shown in Fig \ref{vsepr}. Shown here is $\pm v(r)^{2}$ where $v(r)$ is the
fourier transform of $v(k)$ defined in \cite{kwo95}. 
The sign is chosen to be the sign of $v(r)$.

\begin{figure}
\centerline{
\psfig{figure=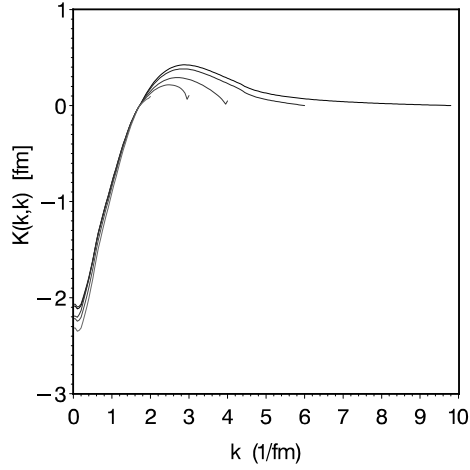,width=7cm,angle=0}
}
\vspace{.0in}
\caption{
The diagonal element of the Brueckner reaction-matrix in the $^{1}S_{0}$
with a starting energy of $-100 Mev$ for different values of $\Lambda$.
}
\label{Kkk}
\end{figure}

Quite a different $\Lambda$-dependence is shown by the Brueckner K-matrix
\begin{equation}
K=V+V{Q\over{e}}K
\label{K}
\end{equation}
that is used to calculate the nuclear matter binding energy. 
Fig \ref{Kkk} shows the
diagonal element $K(\omega=-100MeV,k,k)$. One finds it to be practically
independent of the cut-off. This can to some extent be explained by the
close relationship between the $K$- and the $R$-matrix, the latter being
defined by
\begin{equation}
R=V+VP{1\over{e_{0}}}R
\label{R}
\end{equation}
Here $P$ denotes principal value which in general is not necessary to
include in the definition of the $K$-matrix.
The main difference between the two is in the Pauli-operator $Q$ and the
$"e"$ in the Brueckner reaction-matrix that includes a (selfconsistent)
mean field while $e_{0}$ in the $R$-matrix only has kinetic energies. 
It is important to know that the diagonal elements 
$<k|R|k> \propto tan(\delta(k))/k$ and therefore independent
of the potential within the range of momenta for which the phases are
fitted. For  nuclear matter (or finite nuclei) with densities below 
or slightly above saturation only phases below $k \leq 2 fm^{-1}$ are
required. As pointed out above the $^{1}S_{0}$ potential is well 
appproximated by a separable potentiali. Therefore the
$R$-matrix is also separable and thus completely determined by the
phase-shifts and independent of any potential-representation as long as
the phases are fitted. Because of the relation
\begin{equation}
K=R+RP({Q\over{e}}-{1\over{e_{0}}})K
\label{KR}
\end{equation}
$K$ is also separable and independent of the potential. Because the
separable aproximationis good for $S$-states the off-shell part of the
$K$-matrix needed in eq (\ref{KR}) (or eq (\ref{K}) is also determined by
the phase-shifts to a good approximation.

In a previous paper eq (\ref{KR}) was used for all states not only for the
$^{1}S_{0}$ with surprisingly good results\cite{hsk84}. 
This can also be understood
from the fact that the phase-shift approximation $K\sim R\propto 
tan(\delta(k)/k$
is a good approximation for practically all states except for the two
$S$ and
the $^{3}P_{1}$ states. This will be shown below. 

\begin{figure}
\centerline{
\psfig{figure=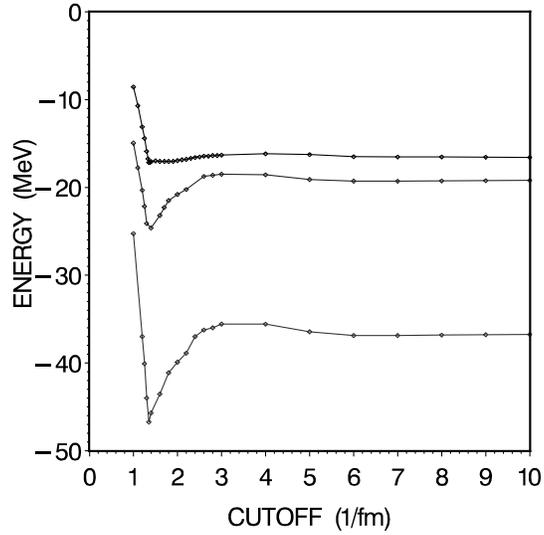,width=8cm,angle=0}
}
\vspace{.0in}
\caption{The three curves show potential energies per particle as a
function of cutoff $\Lambda$ of the phase shifts. The upper curve is
for the $^{1}S_{0}$ state the middle for the $^{3}S_{1}$ and the lowest
includes all the 21 states used. This is for $k_{F}=1.35 fm^{-1}$.
}
\label{vlowfig1}
\end{figure}
\subsection{In-medium Interaction and Nuclear matter}
\begin{figure}
\centerline{
\psfig{figure=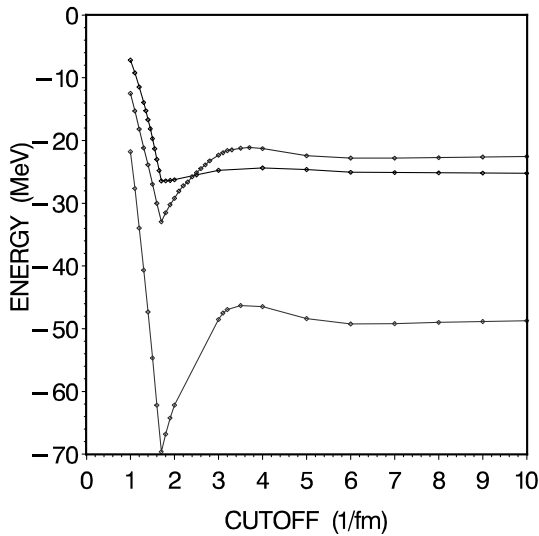,width=8cm,angle=0}
}
\vspace{.0in}
\caption{
Same as Fig \ref{vlowfig1} except for $k_{F}=1.7 fm^{-1}$
}
\label{vlowfig1a}
\end{figure}

\begin{figure}
\centerline{
\psfig{figure=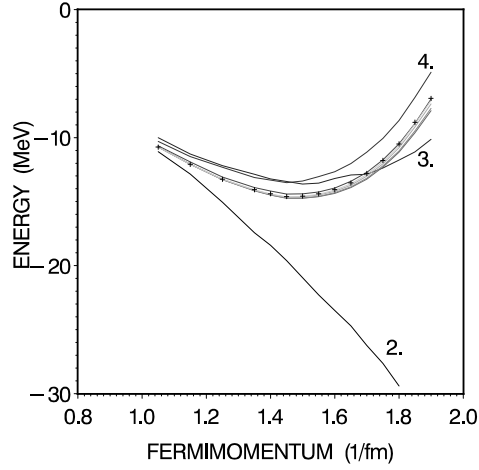,width=7cm,angle=0}
}
\vspace{.0in}
\caption{
The binding energy calculated by standard Brueckner method (and
continuous energy spectrum) as a function of fermi-momentum. Each curve
corresponds to a different cutoff $\Lambda$ of the phase-shifts. For
$\Lambda  \geq 5. fm^{-1}$ all curves nearly coincide. The curve with
crosses is for $\Lambda=10.$ For $\Lambda  \leq 4.
fm^{-1}$ the curves are marked. 
}
\label{vlowfig2}
\end{figure}

\begin{figure}
\centerline{
\psfig{figure=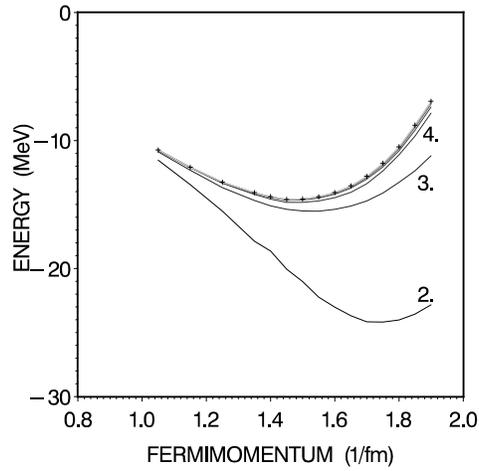,width=7cm,angle=0}
}
\vspace{.0in}
\caption{
Similar to Fig \ref{vlowfig2} but here all phases are cut at
$\Lambda=10. fm^{-1}$ when calculating the potentials but the 
cut-offs indicated in the figure is used when calculating 
the Brueckner $K$-matrix.
}
\label{vlowfig3}
\end{figure}

The main contributions to the binding energy of nuclear matter comes
from the $S$-states. In Fig \ref{vlowfig1} the uppermost curve shows the
contribution to the potential energy per particle
from the $^{1}S_{0}$ state as a
function of the momentum cut-off $\Lambda$ of the phase-shifts. 
For momenta $k > 4.3 fm^{-1}$
a straight line extrapolation to zero at the cut-off is used.
The density is here fixed at
a fermi-momentum $k_{F}=1.35 fm^{-1}$ and the mean-field is that given by
the Brueckner self-consistency at the same $k_{F}$. 
It is seen that the energy here is
almost constant to a cut-off as low as $k_{F}$. 
This is in agreement with Fig \ref{Kkk} showing the near independence of
cut-off for the diagonal elements of
$K$. The only dependence here is for low momenta which have very little
weight in calculating the potentail energy.
For the $^{3}S_{1}$
state (with the Bonn-B deuteron) the situation is different. 
A slight decrease in binding 
is seen already at $5 fm^{-1}$ followed by a sharp increase below $2.5
fm^{-1}$. The lowest curve shows the sum over all states and here the
result for the $^{3}S_{1}$ is even more accentuated.
Fig \ref{vlowfig1a} shows the similar result for $k_{F}=1.7 fm^{-1}$.
It is seen that the lowest momentum accceptable for cut-off has not
shifted appreciably but the potential energys below $5 fm^{-1}$ increase
more drastically than at the lower density.
Fig \ref{vlowfig2} shows the result of binding energy calculations as a
function of density. With cut-offs $\Lambda$
above $5 fm^{-1}$ the curves are well
clustered around a common line but at a cut-off of $4 fm^{-1}$
there is a noticable difference that is in accord with Figs \ref{vlowfig1}
and \ref{vlowfig1a}. Very similar results were reported by J.Kueckei et
al. \cite{kuc02}.
It is of some interest to compare this result with a similar one where the
potentials are calculated with a constant cut-off  $\Lambda= 10 fm^{-1}$ 
while in the
summation over momenta when calculationg the Brueckner K-matrix the
cut-off is varied. The result of such a calculation is shown in Fig
\ref{vlowfig3}. Here a cut-off of $4 fm^{-1}$ is still acceptable but
a cut-off of  $3$  is definitely worse and of course the $\Lambda=2$ cutoff is very
bad.

\begin{figure}
\centerline{
\psfig{figure=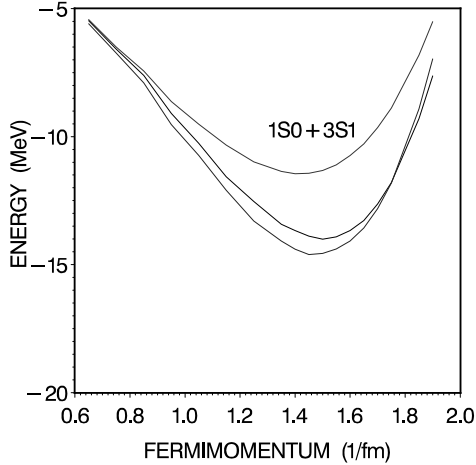,width=7cm,angle=0}
}
\vspace{.0in}
\caption{
The lowest curve shows the binding energy as a function of $k_{F}$ with
all states calculated by
Brueckner while in the center curve only the $^{3}S_{1}$ the $^{1}S_{0}$ and
the $^{3}P_{1}$ are calculated by Brueckner but all other states are by
the phase-shift approximation. The uppermost curve shows the binding
energy with only the $^{1}S_{0}$ and the $^{3}S_{1}$ state 
included.
}
\label{vlowfig5}
\end{figure}

It was shown in an earlier work \cite{hsk84} that the phase-shift
approximation is quite accurate for practically all states beyond the
$S$ even at saturation density. \footnote{ Following \cite{dew56,fuk56} the
approximatiom $\delta /k$ was used there.} 
This is also seen in Fig \ref{vlowfig5}
where in the center curve only the $^{3}S_{1}-^{3}D_{1}$ the $^{1}S_{0}$ and the $^{3}P_{1}$ states
are calculated in Brueckner while the remaining 18 states are in the
phase-shift approximation i.e $\tan(delta)/k$. The phase-shift approximation
fails badly for the three states mentioned but as shown in Fig
\ref{vlowfig5} it is quite accurate for the rest.
It was  also found in ref \cite{hsk84}
that a Pauli and mean
field correction applied to the $S$-states using only the phase-shifts
as input agreed surprisingly well with the standard Brueckner
calculation from a NN-potential.  For these states the many-body
(three-body) effects due to the dispersive corrections stemming from the
short-ranged correlations are very important for saturation as these
corrections are proportional to  
the wound-integral referred to above. \cite{mos60}

\section{Conclusions}
A main goal of nuclear theory is to describe  the properties of nuclei
in terms of the interactions between nucleons. 
The nuclear many-body problem 
has been the subject of intense
investigations and discussions starting with Brueckner's ground breaking
work some 50 years ago. His work showing the importance of Pauli and
dispersion effects to explain the saturation properties of nuclear
matter was ground-breaking. Although much progress has been made 
there remains however still detailed
questions relating both to the underlying interactions between nucleons
and the many-body effects. With the realisation that nucleons are not
elementary particles the concept of a nucleon potential becomes more
obscure except (perhaps) for the long-ranged part.
This will hopefully be circumvented by the EFT methods now being
developped, e.g. in ref.\cite{kol05}

One of the goals of the EFT methods is to develop a low momentum
(long-ranged) interaction  with the short-ranged part due to heavier
exchange particles being integrated out. The purpose of the present
investigation is to see to what extent a traditional approach based
upon fits to phase-shifts depends upon the range of momenta used in the
construction of a potential and subsequently the calculation of nuclear
binding.

The results of this investigation as presented here relates mostly to
the $^{1}S_{0}$ state. This may be a special case because it allows for
the separable representaion to be a good approximation. It is however
found in this case that (at least in the calculation of nuclear binding 
by Brueckner) the potential in itself is irrelevant. The only
requirement is that it fits the phase-shifts for momenta within the region
of interest which does not provide a unique solution of the potential as
it depends quite strongly on the cut-off. But the separable
representaion does provide a practically unique  effective interaction 
($K$-matrix) independent of cut-off.

The $^{3}S_{1}-^{3}D_{1}$ states will be the subject of  another paper.
Here the biggest source of uncertainty relates to the deuteron $D$-state
probability $P_{D}$.
The results shown in Figs \ref {vlowfig1} and \ref{vlowfig1a}
do however indicate a much greater sensitivity to cut-offs below $4-5
fm^{-1}$.

It was shown in Fig \ref{vlowfig5} that for  higher angular momentum
states (except the $^{3}P_{1}$) the phase-shift approximation is quite
adequate. 

A topic of interest is to what extent the experimentally undetermined
off-shell part of the potential affects the in-medium $K$-matrix which
because of the propagation through a mean-field is off-shell. In the
results presented here this question is not answered. The off-shell
$K$-matrix is here constrained by the choice of lowest possible rank of
the interaction. It was one the incentives of this work that  a change
of the off-shell behavior is possible by increasing the rank of 
the potential while
keeping the on-shell data intact.  The above-mentioned question could
then be
answered. This would be very difficult by traditional potential
construction using meson-theoretical input, but in principle easy within
the separable method implemented here.

That the off-shell behavior does indeed affect the in-medium interaction is
exemplified by the $^{3}P_{1}$ case which deviates
appreciably from the Bonn-result. In the previous work with the
separable potential the comparison with the Bonn-B potential half-shell
reactance matrix was shown to be very different from that of the
separable potential.\cite{kwo95}
The rank-1 separable potential used in this case
appears not to be appropriate.

Acknowledgement:
Helpful discussions with Professor Nai Kwong are greatly appreciated.

\end{document}